# Application of Zeeman spatial beam-splitting in polarized neutron reflectometry

S.V. Kozhevnikov[1*], V.K. Ignatovich[1], F. Radu[2]

[1]*Frank Laboratory of Neutron Physics, JINR, 141980 Dubna, Russian Federation*
[2]*Helmholtz-Zentrum Berlin für Materialien und Energie, Albert-Einstein Straße 15, D-12489 Berlin, Germany*

Neutron Zeeman spatial beam-splitting is considered at reflection from magnetically noncollinear films. Two applications of Zeeman beam-splitting phenomenon in polarized neutron reflectometry are discussed. One is the construction of polarizing devices with high polarizing efficiency. Another one is the investigations of magnetically noncollinear films with low spin-flip probability. Experimental results are presented for illustration.

## I. INTRODUCTION

Polarized neutron reflectometry (PNR) is by now a mature method in the field of magnetic heterostructures. It is used, in particular, for investigations of thin magnetic films and multilayers [1] for measurements of magnetization distribution along the normal to the sample plane. The conventional scheme of a polarized neutron reflectometer consists of polarizer, the first spin-flipper, investigated sample, the second spin-flipper, analyzer and detector. The final neutron intensity registered by detector contains the sample reflectivity (which is the useful signal) and imperfection of polarizing devices (which is the parasitic background). Thus, to extract lower useful signal from the magnetic film, we have to reduce parasitic background from polarizing devices. Therefore the increasing of polarizing devices efficiency is an actual task.

One of such tool can be the Zeeman spatial splitting of the neutron beam at reflection from a magnetically noncollinear film. If spin-flip takes place in a high magnetic field, neutron beams of different spin-flip transitions are separated in space in different off-specular regions. At the same time, non spin-flipped neutrons are reflected in specular reflection region. Thus, Zeeman beam-splitting directly extracts in space definite useful spin-flip signal from other ones and consequently reduces parasitic background. This property can be exploited in two ways. One is creation of a polarizer with high polarizing efficiency.

Another one is using the Zeeman beam-splitting to investigate the magnetically noncollinear film itself. In this communication we describe the method of Zeeman beam-splitting and illustrate two mentioned ways of its application by experimental data.

## II. THE ZEEMAN BEAM-SPLITTING

The Zeeman spatial splitting of the neutron beam takes place at reflection and refraction at boundary of two magnetically noncollinear media. This phenomenon was predicted theoretically in [2] and observed experimentally in the geometry of reflection in [3-5] and refraction in [6-9]. The beam-splitting was also registered at reflection from thin magnetically anisotropic films with domains [10-12], from internally anisotropic super-lattices [13,14] and clusters [15,16]. The beam-splitting phenomenon was applied to direct determination of the magnetic induction in magnetically noncollinear media [17-19] and investigations of magnetically noncollinear media themselves.

Here we briefly consider the geometry of the beam-splitting experiments. More detailed description of the beam-splitting principle and data representation were given in [17]. In Fig. 1, the geometry of an experiment for the neutron reflection from and transmission through a magnetic film with induction $B$ sputtered on a nonmagnetic substrate with nuclear potential $U$ put in a magnetic field $H \approx 10$ kOe applied in $(y,z)$ plane under an angle $\alpha$ to the sample surface is presented. The sample surface is $(x,y)$ plane. $Oz$ axis is perpendicular to the sample surface. $Ox$ axis is parallel to the sample surface. The incident polarized beam with spin (+)/along or (-)/opposite to the external field falls under the grazing angle $\theta_0$. The final angle of reflected or refracted beam is $\theta$. The specular reflection takes place at $\theta=\theta_0$ and off-specular reflections correspond to $\theta \neq \theta_0$. The spin-flip probability $W \sim \sin^2 \chi$ depends on the angle $\chi$ between vectors of the external magnetic field **H** and internal magnetic induction **B** (see [2,20]). It was measured experimentally in [6,21].

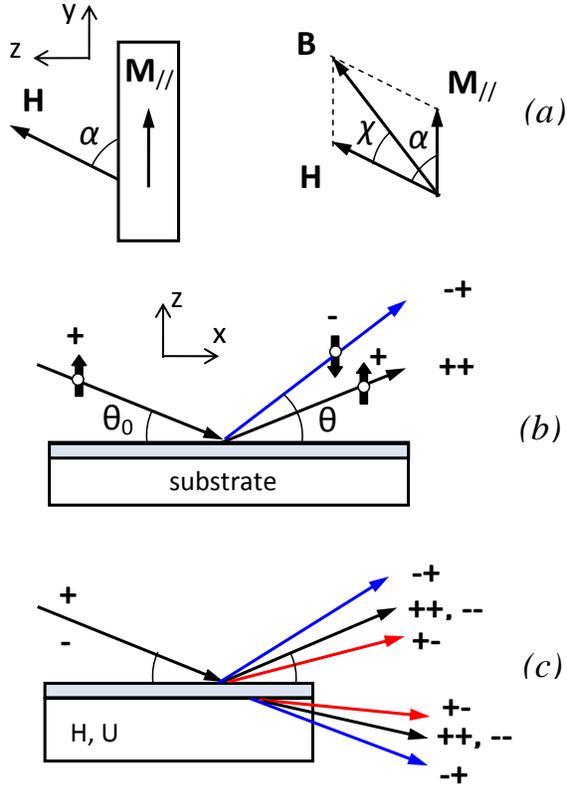

Fig. 1. Geometry of the neutron beam-splitting experiment. *(a)* Directions of outside, **H**, and inside, **B**, magnetic fields. The spin-flip probability, $W$, is proportional to $\sin^2\chi$ of the angle $\chi$ between **H** and **B**. *(b)* The beam-splitting at reflection for the incident beam (+), polarised along **H**. *(c)* The beam-splitting at reflection and refraction of the nonpolarized incident neutron beam having both components: polarized along, (+), and opposite, (-), the external field **H**.

Here we consider only reflection. The component of the wave vector of the incident neutron parallel to the interface $k_{0x} = k_0 \cos\theta_0$ does not change at reflection, the normal component $k_{0z} = k_0 \sin\theta_0$ also does not change in the case of specular reflection without spin-flip, but it changes to $k_{\pm z} = \sqrt{k_{0z}^2 \pm 2u_H}$, when the spin direction changes. Here $u_H = 2m|\mu|H/\hbar^2$ is the neutron magnetic interaction, m, μ are the neutron mass and magnetic moment respectively. The component $k_{+z}$ corresponds to reflection with spin-flip from the initial state (+) along the external field **H** to the final state (-) opposite to **H**. In Fig-s 1*b* and 1*c* such a component is denoted by the symbol (-+). The component $k_{-z}$ corresponds to reflection with spin-flip from the initial state (-) opposite to the external field **H** to the final state (+) along **H**. In Fig. 1*c* such a component is denoted by the symbol (+-). The change $\Delta\theta_\pm$ of the angle after reflection can be found from the relation

$$\sqrt{k_0^2 \pm 2u_H}\, \sin(\theta_0 \pm \Delta\theta_\pm) = \sqrt{k_{0z}^2 \pm 2u_H} \quad (1)$$

In the case of thermal incident neutrons with wavelength λ=1.8Å the angles $\theta_0$ and $\theta_\pm = \theta_0 \pm \Delta\theta_\pm$ are of the order of $10^{-2}$, therefore Eq. (1) can be represented as

$$\theta_0 \pm \Delta\theta_\pm = \sqrt{\theta_0^2 \pm \alpha}\Big/\sqrt{1 \pm \alpha} \quad (2)$$

where $\alpha = 2u_H/k_0^2$. Since for thermal neutrons $\alpha \propto 10^{-5}$ then Eq. (2) can be represented as

$$\pm \Delta\theta_\pm \approx \sqrt{\theta_0^2 \pm \alpha} - \theta_0 \quad (3)$$

If $\theta_0^2 < \alpha$, then $\Delta\theta_- = \theta_0$, or $\theta_- = 0$, which means that the beam (+-) does not appear at these low angles $\theta_0$.

### III. EXPERIMENTAL SETUP

The experiments were done at the polarized neutron TOF reflectometer SPN-2 at the pulsed reactor IBR-2 (Frank Laboratory of Neutron Physics, Joint Institute for Nuclear Research, Dubna, Russia). The sample surface plane *(x,y)* is vertical and the beam scattering plane *(x,z)* is horizontal. The experimental setup is shown in Fig. 2*a*. The reactor pulse is the start moment for measurement of the neutron flight time. The thermal neutrons after moderator (M) are polarized in the curved 5 m long FeCo polarizer (P). The curvature radius of the polarizer 1 km defines the characteristic minimal wavelength 1.0 Å. The cross section of the polarized beam at the exit of the polarizer was 2.5(horizontal)×60(vertical) mm². The polarization of the beam before the sample was reversed by non-adiabatic spin-flipper (SF1) of Korneev type [23,24]. The sample (S) was placed between the poles of the electromagnet (EM). The magnetic field can be rotated in the plane perpendicular to the sample surface in the interval 0-90°. The second adiabatic radiofrequency spin-flipper (SF2) [25] had the diameter of 100 mm and was used to reverse the scattered beam polarization. The polarization of the scattered beam was analyzed by the multislit curved supermirror analyzer (A) with 32 horizontal mirrors of 10 m curvature radius. The sizes of the mirrors were 300×50×0.2 mm$^3$. The distance between the stacked mirrors was 1 mm. The critical wavelength for this curved analyzer was 1.8 Å. The working area of the analyzer was 38×40 mm$^2$. The neutron beam was registered by one-coordinate $^3$He position-sensitive detector (PSD) with the working area 120 (horizontally) × 40 (vertically) mm$^2$ and spatial resolution 1.5 mm [26]. The time between the reactor pulse and the moment of registering of neutrons by the detector is the time of neutron flight along the distance between moderator and the detector which was 32-37 meters. The distance 'moderator-sample' was fixed at 29 m. The distance 'sample-detector' was 3 or 8 m. The width of the reactor pulse was 320 μs. So the neutron wavelength resolution was 0.02 Å for the TOF base 37 m.

The polarization efficiencies of the polarizer and the analyzer were defined by 3P2S (3 polarizers and 2 spin-flippers) method [27-29]. In Fig. 2b, efficiencies of polarizer (squares) and analyzer (circles) are presented in dependence on neutron wavelength. Efficiency of the analyzer is rather high in almost the whole spectrum. But the polarizer efficiency significantly decreases at long wavelengths. This fact is explained by reflection of long wavelength neutrons from the absorbing TiGd layer [30], which increases contamination of the polarized beam by neutrons with the opposite polarization. Some parameters of the spectrometer SPN can be found in [27].

## IV. IMPROVEMENT OF POLARIZING EFFICIENCY

In an experiment with polarizer, spin flippers before and after sample, and analyzer one measures detector counts at different settings, on or off, of the flippers. The count rate of the detector for an unpolarized primary beam can be represented as [27]

$$I^{j,i} = \langle 0|\mathbf{AF_2^j R F_1^i P}|0\rangle I_0/2 \qquad (4)$$

where $I_0$ is the intensity of the incident beam, indices $i, j$ mean the flippers state on or off, **A, F, P** denote matrices describing actions of analyser, spin-flipper and polarizer respectively

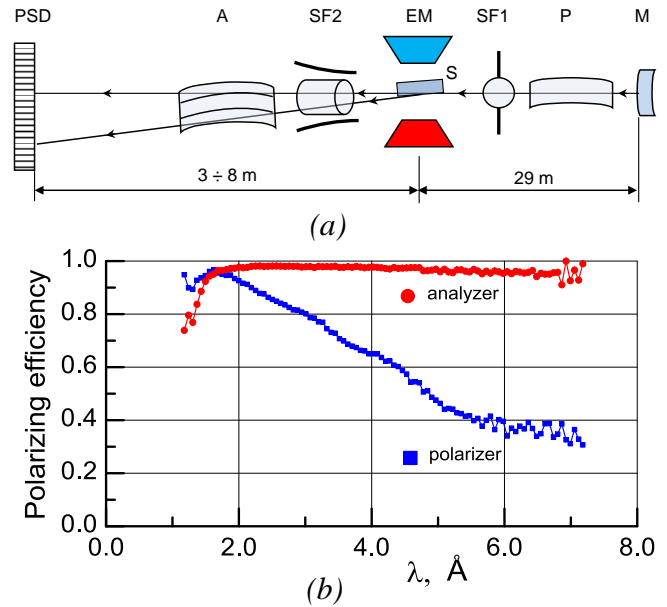

Fig. 2. *(a)* Scheme of the polarized neutron TOF reflectometer SPN at the pulsed reactor IBR-2: M is the water moderator, P is a polarizer which is a 5 m long curved neutron guide with vertical slit shape cross section, SF1 is the Korneev non-adiabatic spin-flipper, S is the sample with vertically oriented surface, EM is electromagnet, SF2 is the adiabatic radiofrequency spin-flipper for a scattered beam which is scattered in the horizontal plane, A is the multislit curved supermirror analyser with horizontally oriented mirrors, PSD is one-coordinate $^3$He position-sensitive detector oriented horizontally. *(b)* Polarizing efficiencies of the polarizer (squares) and the analyzer (circles) as a function of neutron wavelength. Polarizing efficiency of the polarizer decreases at long wavelength because the absorbing layer starts to reflect the neutrons with an opposite parasitic spin.

$$\mathbf{A} = \begin{pmatrix} 1+P_a & 0 \\ 0 & 1-P_a \end{pmatrix}, \qquad \mathbf{F} = \begin{pmatrix} 1-f & f \\ f & 1-f \end{pmatrix},$$
$$\mathbf{P} = \begin{pmatrix} 1+P_p & 0 \\ 0 & 1-P_p \end{pmatrix} \qquad (5)$$

where $P_a$, $P_p$ are polarizing efficiencies of analyser and polarizer respectively, $2f-1$ is efficiency of the spin-flipper, when it is in "on" state. In the "off" state the parameter $f$ is zero. Matrix **R** denotes reflectivity matrix of the sample. It has matrix elements $R^{j,i}$, where $j,i$ is + or -, which are equal to reflection probability with or without spin flip. The right index denotes initial, and left one – the final states. The brackets $\langle 0|=(1,1)$ and $|0\rangle = \langle 0|^T$ denote classical two dimensional bra and ket-vectors, respectively.

Relation (4) gives four equations. One of them for both flippers in the state "on" is

$$\frac{I_0}{2}(1\ 1)\begin{pmatrix}1+P_a & 0\\ 0 & 1-P_a\end{pmatrix}\begin{pmatrix}1-f_2 & f_2\\ f_2 & 1-f_2\end{pmatrix}\begin{pmatrix}R^{++} & R^{+-}\\ R^{-+} & R^{--}\end{pmatrix}\begin{pmatrix}1-f_1 & f_1\\ f_1 & 1-f_1\end{pmatrix}\begin{pmatrix}1+P_p & 0\\ 0 & 1-P_p\end{pmatrix}\begin{pmatrix}1\\1\end{pmatrix}=$$

$$=(I_0/2)\{(1-P_{fa})[(1-P_{fp})R^{++}+(1+P_{fp})R^{+-}]+(1+P_{fa})[(1-P_{fp})R^{-+}+(1+P_{fp})R^{--}]\}=I^{on,on} \qquad (6)$$

where $P_{fa}=P_a(2f_2-1)$ and $P_{fp}=P_p(2f_1-1)$. The other three are obtained from it by putting one of $f_i$ or both to zero. The system of these equations is equal to

$$\frac{I_0}{2}\begin{pmatrix}(1-P_{fa})(1-P_{fp}) & (1-P_{fa})(1+P_{fp}) & (1+P_{fa})(1-P_{fp}) & (1+P_{fa})(1+P_{fp})\\ (1-P_{fa})(1+P_p) & (1-P_{fa})(1-P_p) & (1+P_{fa})(1+P_p) & (1+P_{fa})(1-P_p)\\ (1+P_a)(1-P_{fp}) & (1+P_a)(1+P_{fp}) & (1-P_a)(1-P_{fp}) & (1-P_a)(1+P_{fp})\\ (1+P_a)(1+P_p) & (1+P_a)(1-P_p) & (1-P_a)(1+P_p) & (1-P_a)(1-P_p)\end{pmatrix}\begin{pmatrix}R^{++}\\R^{+-}\\R^{-+}\\R^{--}\end{pmatrix}=\begin{pmatrix}I^{on,on}\\I^{on,off}\\I^{off,on}\\I^{off,off}\end{pmatrix} \qquad (7)$$

So the four reflectivities $R^{++}$, $R^{+-}$, $R^{-+}$ and $R^{--}$ from the magnetically noncollinear film can be extracted from the four intensities of the reflected beams [27-29] $I^{off,off}$, $I^{off,on}$, $I^{on,off}$ and $I^{on,on}$ registered by the detector, if the detector overlaps all the reflected beams with and without spin-flip.

When the sample is absent, the matrix **R** in (4) can be replaced by the unit matrix, and (7), after substitution $R^{++}=R^{--}=1$, $R^{-+}=R^{+-}=0$ gives the system of equations

$$I_0(1+P_{fa}P_{fp})=I^{on,on},\quad I_0(1-P_{fa}P_p)=I^{on,off},\quad I_0(1+P_aP_p)=I^{off,off},\quad I_0(1-P_aP_{fp})=I^{on,off} \qquad (8)$$

for determination of the four parameters $f_1$, $f_2$, $P_pP_a$ and $I_0$.

To separate the polarizing efficiencies $P_p$ and $P_a$, a magnetically collinear saturated mirror as a calibrator is used: $R_C^{++}\neq R_C^{--}\neq 0$ and $R_C^{-+}=R_C^{+-}=0$. We put these reflectivities and the defined from (8) values $f_1$ and $f_2$ to (7) and extract the parameters $P_p$ and $P_a$. To define the reflectivities $R_C^{++}$ and $R_C^{--}$ we have to put also the obtained from (8) parameter $I_0$ to (7).

To simplify further analysis, we suppose $f_1=f_2=1$ and $P_a=1$. Then $P_{fa}=1$ $P_{fp}=P_p$, and Eq. (7) becomes

$$I_0\begin{pmatrix}0 & 0 & (1-P_p) & (1+P_p)\\ 0 & 0 & (1+P_p) & (1-P_p)\\ (1-P_p) & (1+P_p) & 0 & 0\\ (1+P_p) & (1-P_p) & 0 & 0\end{pmatrix}\begin{pmatrix}R^{++}\\R^{+-}\\R^{-+}\\R^{--}\end{pmatrix}=\begin{pmatrix}I^{on,on}\\I^{on,off}\\I^{off,on}\\I^{off,off}\end{pmatrix} \qquad (9)$$

So the four equations split into two systems of two equations, and their solutions for, say, spin-flip reflectivities are

$$R^{+-}=\frac{1}{8P_p}\left[I^{off,on}(P_p+1)+I^{off,off}(P_p-1)\right] \qquad (10)$$

$$R^{-+}=\frac{1}{8P_p}\left[I^{on,off}(P_p+1)+I^{on,on}(P_p-1)\right] \qquad (11)$$

For reflectivity $R^{-+}$ contribution $I^{on,off}(P_p+1)$ is the useful signal, and the part $I^{on,on}(P_p-1)$ is the background. To decrease the statistical error one can improve the polarizing efficiency $P_p$ up to 1. Another way is to suppress the background intensity $I^{on,on}$ by using the off-specular region for polarization analysis. We illustrate this approach by experimental data.

In Fig. 3 the results of measurement in specular (Fig. 3*a*) and nonspecular (Fig. 3*b*) regions are presented, when the external field 3.5 kOe is parallel to the sample surface and therefore is collinear to the internal field. In Fig. 3*a*, the reflectivity '+-' is shifted on -0.5 for clarity. In Fig. 3*b*, the reflectivity '++' is shifted on +1.0, '--' is shifted on +0.5 and '+-' is shifted on -0.5. In such geometry there should be no spin-flip reflectivities and no nonspecular counts. The sample was a thin film Co(700 nm)/glass(substrate) with the sizes 100(along the beam)×50(width)×5(substrate thickness) mm$^3$. The glancing angle of the incident beam was $\theta_0 = 0.21°$. It is clearly seen, that in specular region $R^{+-} = R^{-+} = 0$, and in the off-specular region there are no counts at all, i.e. $R^{++} = R^{--} = R^{+-} = R^{-+} = 0$. In Fig. 4 the error bar $\delta R^{-+}$ for the reflectivity $R^{-+}$ in specular and off-specular regions is shown as a function of the neutron wavelength. One can see that in the interval of neutron wavelength 2.5 - 6.0 Å the error bar in the off-specular region is 10 times smaller than in the specular one. Thus, the background in the off-specular region is suppressed.

When the external field is applied at an angle $\alpha = 80°$ to the sample surface, the external and internal fields become noncollinear, and the off-specular reflections with spin-flip do appear. It is seen in the two-dimensional map of the neutron intensity presented in Fig. 5 as a function of the neutron wavelength and the glancing angle of the scattered beam. The glancing angle of the incident beam was $\theta_0 = 0.21°$. The solid horizontal lines at the angles +0.21°, 0° and -0.21° correspond to the specularly reflected beam, the sample horizon and the direct beam direction, respectively. In the case of low external field 220 Oe there is practically no beam-splitting in the modes *off,on* (Fig. 5*a*) and *on,off* (Fig. 5*b*). In the high field 3.5 kOe the beam-splitting in these modes *off,on* (Fig. 5*c*) and *on,off* (Fig. 5*d*) is clearly seen.

In Fig. 6 the reflected beam intensity integrated over the neutron wavelength interval 2.55 ÷ 6.67 Å is

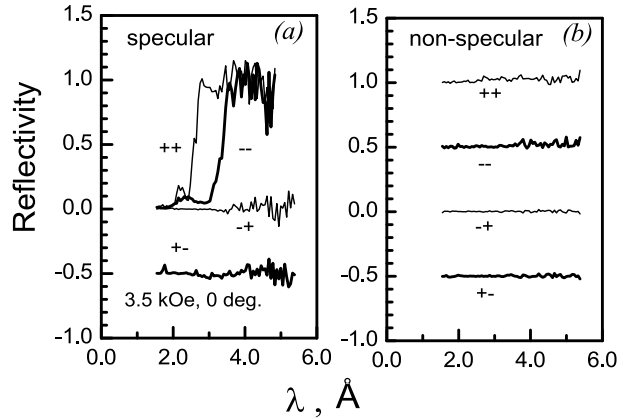

Fig. 3. Experimental reflectivities as a function of the neutron wavelength for the Co(70nm)//glass film in the field 3.5 kOe applied parallel to the sample surface.
*(a)* Specular reflection '++', '--', '-+' and '+-' (shift -0.5 for clarity). *(b)* Off-specular reflection '++' (shift +1.0), '--' (shift +0.5), '-+' and '+-' (shift -0.5).

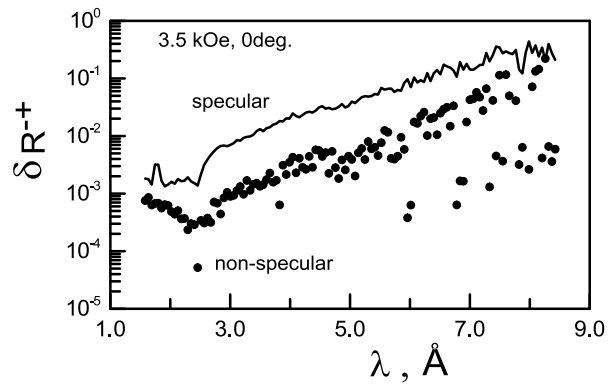

Fig. 4. The error bar of the reflectivity '-+' in specular (line) and off-specular (symbol) regions in the field 3.5kOe applied parallel to the sample Co(70nm)//glass surface.

presented as a function of the glancing angle of the reflected beam. The glancing angle of the incident beam was $\theta_0 = 0.21°$. In Fig.6*a* the angular distributions are shown for collinear strong field 3.5 kOe and noncollinear weak field 220 Oe. In Fig.6*b* the angular distributions for the noncollinear strong field 3.5 kOe are shown. In the collinear geometry there is no spin-flip therefore intensity in the mode *off,on* is only a background for spin-flip. In the low field 220 Oe (Fig. 6*a*) the angular splitting is not resolvable and the ratio *effect/background* is equal to 1.8 at the peaks maxima in the specular reflection region. In the high magnetic field (Fig. 6*b*) in off-specular region the maximal ratio *effect/background* is between 10 and 20. Thus, the ratio *effect/background* in the off-specular

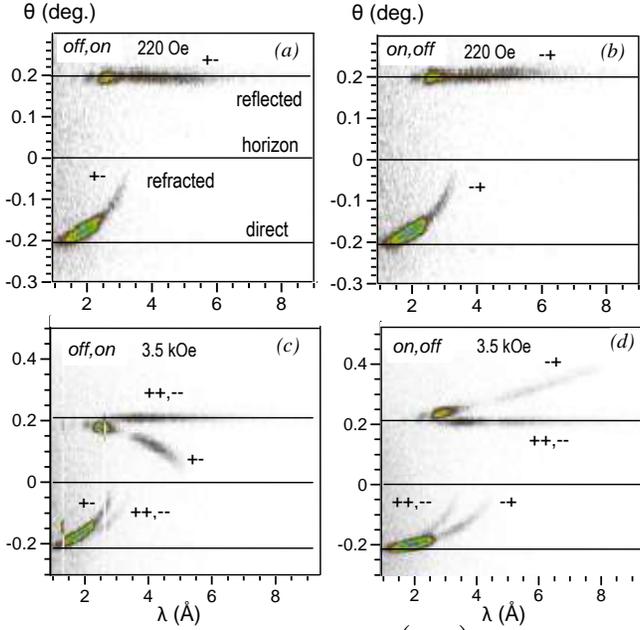

Fig. 5. Two-dimensional map $(\lambda, \theta)$ of neutron spin-flip intensity: upper row *(a)*, *(b)* is the applied field 220 Oe; bottom row *(c)*, *(d)* is the applied field 3.5 kOe; left column *(a)*, *(c)* is *off,on* mode; right column is *on,off* mode. The external field is applied under an angle 80° to the sample Co(70nm)//glass surface.

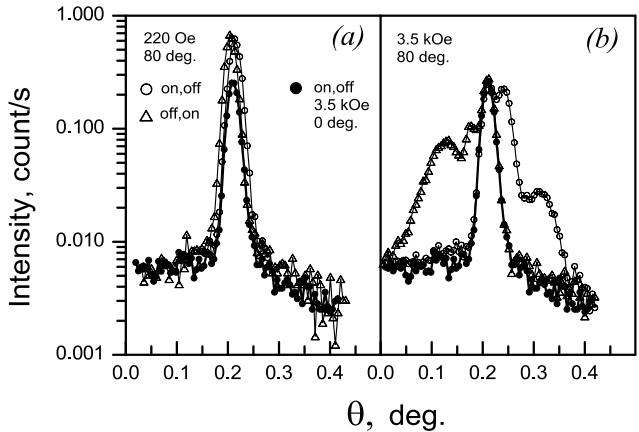

Fig. 6. The neutron intensity as a function of the scattered beam angle: *(a)* in the perpendicular field 220 Oe; *(b)* in the perpendicular field 3.5 kOe. The open circle is *on,off* mode and the open triangle is *off,on* mode. The closed circle is the intensity *on,off* in the parallel field 3.5 kOe (this intensity can be considered as the background for spin-flip).

region is enhanced 10 times comparing to the specular reflection region.

In Fig. 7a,b the spin-flip reflectivities in off-specular region at the high field 3.5 kOe applied at an angle 80°

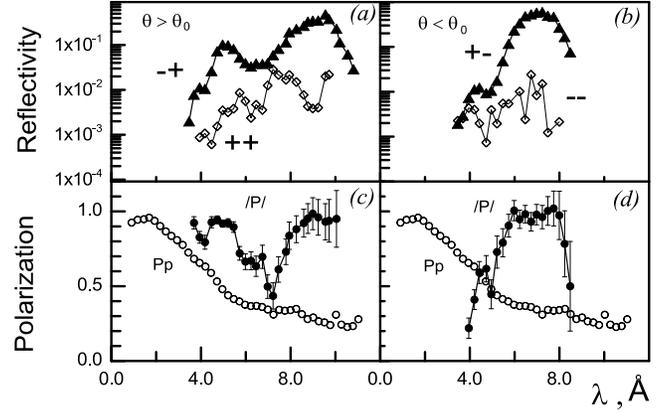

Fig. 7. Spin-flip reflectivity (upper row) and polarization degree of the neutrons beam (bottom row) in off-specular regions $\theta > \theta_0$ (left column) and $\theta < \theta_0$ (right column). Open circles correspond to the polarizing efficiency of the polarizer.

to the sample surface is shown as a function of the neutron wavelength. The glancing angle of the incident beam is $\theta_0 = 0.35°$. Fig. 7a corresponds to the spin-flip '-+' in the region $\theta > \theta_0$ and Fig. 7b corresponds to the spin-flip '+-' in the region $\theta < \theta_0$. The polarization degree of the reflected beam is defined by the expression:

$$P = \frac{R^{++} + R^{+-} - R^{--} - R^{-+}}{R^{++} + R^{-+} + R^{--} + R^{+-}} \quad (10)$$

In the off-specular region $\theta > \theta_0$ (Fig. 7c) we have $R^{-+} \neq 0$, $R^{++} \approx R^{--} \approx R^{+-} \approx 0$ and $P \approx -1$. In the interval of neutron wavelength $8 \div 10$ Å the averaged absolute polarization degree is $|P| = 0.94 \pm 0.11$. In the off-specular region $\theta < \theta_0$ (Fig. 7d) we have $R^{+-} \neq 0$, $R^{++} \approx R^{--} \approx R^{-+} \approx 0$ and $P \approx +1$. In the interval of neutron wavelength $6 \div 8$ Å the averaged polarization degree is $P = 0.97 \pm 0.08$. One can see that polarization degree of the spin-flip reflected neutrons in off-specular regions for the long wavelength neutrons is close to 1. And the polarization of the incident neutron beam for the wavelength $\lambda > 5$ Å is $P_p = 0.3$. It means that the beam-splitting effect improves polarizing efficiency of polarizer.

To estimate the quality of polarizers for polarized neutron experiment, the factor $P^2I$ or $P^2R$ must be calculated, where $P$ is polarizing efficiency, $I$ is neutron intensity after polarizer and $R$ is reflectivity of polarizer. The minimal measuring time for the same error bar of the sample reflectivities is achieved at the maximum of the factor $P^2I$ or $P^2R$. The conventional polarizers based on supermirrors have reflectivity about $R=0.95$. The polarizing efficiency is decreasing with increasing neutron wavelength. It deals with worse absorption of '-' component of the neutron intensity inside the absorbing layer in supermirror. For example, polarizing efficiency of fan multislit analyzer [31] is equal to 0.95 for the neutron wavelength 3 Å, 0.90 for 5 Å, 0.85 for 7 Å and 0.80 for 8 - 10 Å. For example, for the polarized beam (+-) in Fig. 7b at the neutron wavelength 7.2 Å we can see $R=0.6$ and in Fig. 7d we see $P=0.97$. Thus, the factor is $P^2R = 0.56$. Corresponding factor for supermirror for 7 Å is $P^2R = 0.85^2 \cdot 0.95 = 0.68$. For the beam (-+) for the neutron wavelength 9.5 Å we have $R=0.53$ (Fig. 7a) and $P=0.94$ (Fig. 7c). The factor is $P^2R = 0.94^2 \cdot 0.53 = 0.47$. Corresponding factor for 9.5 Å for the supermirror is $P^2R = 0.80^2 \cdot 0.95 = 0.61$.

One can see that the factor $P^2R$ for the supermirror is greater than for beam-splitting but not much for the long neutron wavelength. In the case of low spin-flip probability when spin-flip signal is close to background, the higher polarizing efficiency may be more important than the low reflectivity. In this case the beam-splitting effect may be more efficient than supermirror polarizer. Also the polarizer based on the beam-splitting can be used as monochromator with adjustable wavelength band [32].

## V. NEUTRON STANDING WAVES

In this section application of the beam-splitting effect for physical investigations is demonstrated. In particular it can be used for observation of the neutron standing waves using spin-flip in magnetically noncollinear layer [33]. Neutron standing waves for the investigation of magnetic films using spin-flip were used in [34]. The principle of neutron standing waves is shown schematically in Fig. 8.

The sample is the thin film Ti(30nm)/Co(6)/Ti(200)/Cu(100)//glass(substrate). The optical potential of the neutron interaction with matter is of well-like type (Fig. 8a). Neutrons tunnel through the upper thin Co layer and are reflected from the bottom Cu layer. In the middle Ti (200 nm) layer the neutron wavefunction density is resonantly enhanced because of multiple reflection of the neutron wave from top and bottom layers and the resonant phase conditions

$$\gamma(k_{0z}) = 2k_{2z}d + \arg(R_{21}) + \arg(R_{23}) = 2\pi n \quad (11)$$

where $k_{2z}$ is the normal (along $z$-axis) to the sample surface component of the neutron wave vector inside the resonant layer Ti (200 nm); $d=200$ nm is the thickness of the resonant Ti layer; $R_{21}$ and $R_{23}$ are the neutron reflection amplitudes inside the Ti layer from the top (tunneling) Co layer and the bottom Cu layer (reflector) respectively; n=0, 1, 2, ... are the orders of the resonances.

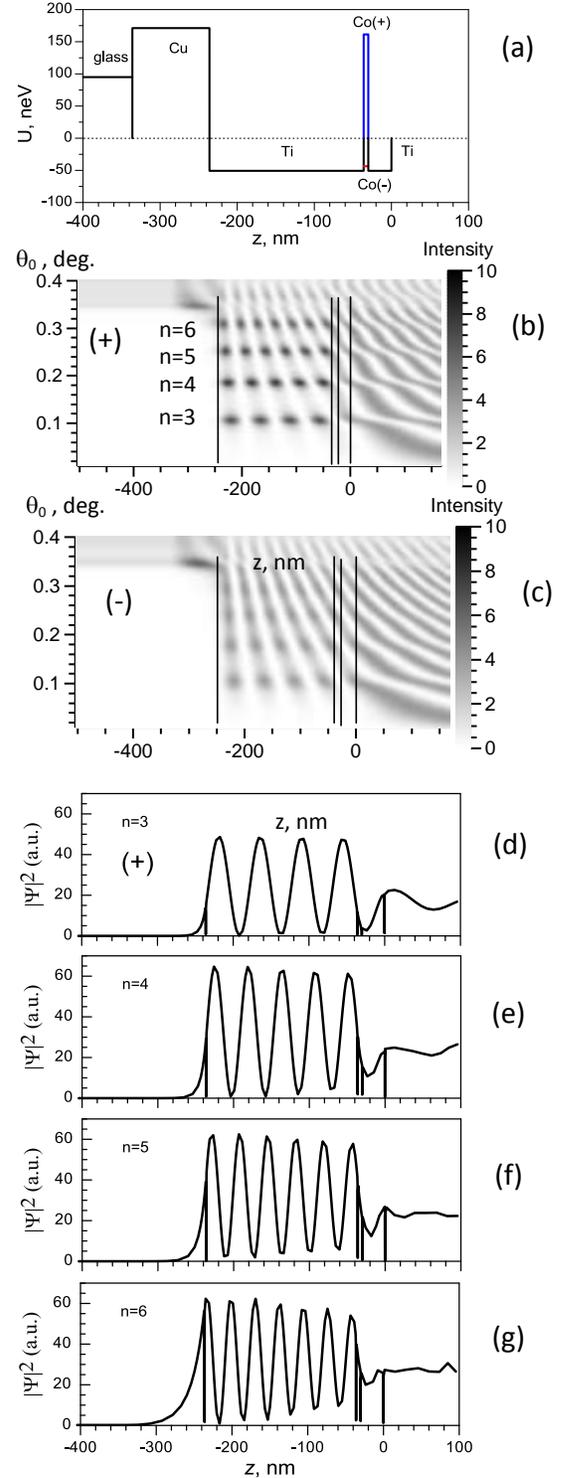

Fig. 8. Calculations for the neutron standing waves for the neutron wavelength 4.0 Å. *(a)* Neutron optical potential of the sample Ti(30 nm)/Co(6 nm)/Ti(200 nm)/Cu(100 nm)//glass(substrate). Neutron wavefunction density as a function of the glancing angle of the incident beam and the coordinate z perpendicular to the sample surface: *(b)* spin (+); *(c)* spin (-). Neutron wavefunction density for the different orders of the resonances n as function of the depth into the sample: *(d)* n=3; *(e)* n=4; *(f)* n=5; *(g)* n=6.

In Fig. 8b and 8c the calculated neutron wavefunction densities along z coordinate for different glancing angles of the incident neutrons with spin (+) and (-) respectively are presented. The neutron wavelength is fixed at 4 Å. In Fig. 8d-g the neutron wavefunction densities in the resonator for the incident neutron spin (+) is shown in dependence on z for the resonance orders n=3, 4, 5 and 6 respectively. One can see 4, 5, 6 and 7 maxima respectively.

The integrated over distance z wavefunction densities for the incident neutrons with spin (+) and (-) are presented in Fig. 9a in dependence on the glancing angle of the incident beam. The maxima corresponding to the resonances n=3, 4, 5 and 6 are clearly seen. The neutron wavefunction density for spin (+) is rather higher than for spin (-). The reason is that the optical potential of the Co layer for the spin (+) is higher than for the spin (-) as shown in Fig. 8a.

Calculated neutron reflectivities for the external magnetic field 150 Oe parallel to the sample surface are presented in Fig. 9b. In this case there are no spin-flip reflectivities. The minima on the total reflection plateau of the non spin-flip reflectivities $R^{++}$ and $R^{--}$ appear because of absorption in the resonant layer, which increases, when standing waves are formed in it. In Fig. 9c calculated reflectivities $R^{++}$ and $R^{-+}$ for the external magnetic field 150 Oe applied under an angle $80^0$ to the sample surface are presented. Minima of non spin-flip amplitude appear because of the spin-flip processes, which are enhanced, when standing waves are formed. They correspond to maxima of the spin-flip reflectivity.

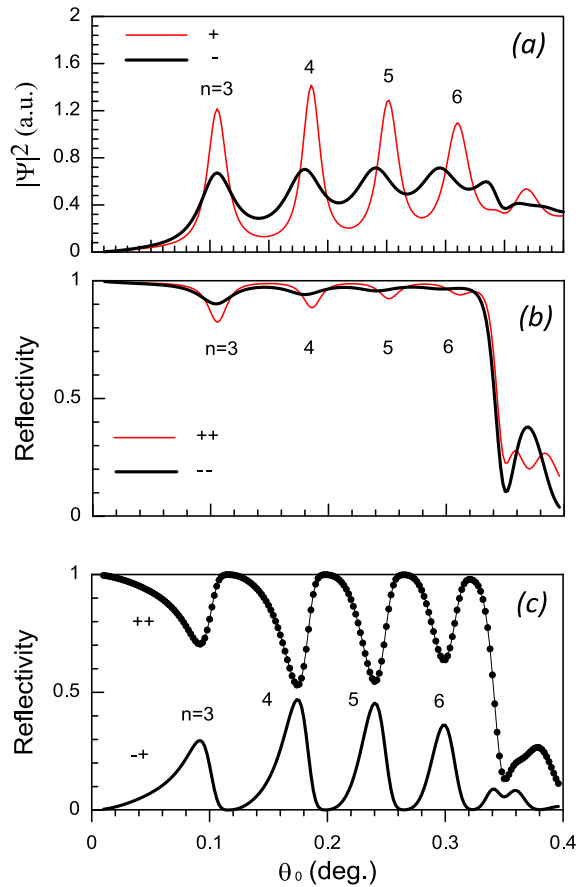

Fig. 9. Calculations for the neutron standing waves for the neutron wavelength 4.0 Å in the sample Ti(30 nm)/Co(6 nm)/Ti(200 nm)/Cu(100 nm)//glass(substrate). *(a)* Neutron wavefunction density for spin (+) and (-) as a function of the glancing angle of the incident beam. *(b)* Non spin-flip reflectivities (++) and (--) as a function of the glancing angle of the incident beam for the field 150 Oe applied parallel to the sample surface. *(c)* Non spin-flip (++) and spin-flip (-+) reflectivities as a function of the glancing angle of the incident beam for the field 150 Oe applied under an angle 80° to the sample surface.

The experiment had been carried out at the polarized neutron reflectometer SPN-2. The sample Ti(30 nm)/Co(6)/Ti(200 )/Cu(100)//glass (substrate) had the sizes 100×50×5 mm$^3$. The external magnetic field was applied under an angle 80° to the sample surface to create magnetic non-collinearity to provide spin-flip processes. The distance sample-detector was 3 m, the glancing angle of the incident beam was $\theta_0 = 0.18°$ with the divergence $\delta\theta_0 = 0.05°$. More experimental details can be found in [33]. In Fig. 10 the neutron intensity at the applied field 6.75 kOe (Figs. 10a,b) in off-specular reflection region and 150 Oe (Figs. 10c,d) in specular reflection region for the state on,off (Figs. 10a,c) and off,on (Fig. 10b,d) are presented as a function of the neutron wavelength. Open symbols corresponds to the inclined applied field. Closed symbols correspond to the magnetic field 4.6 kOe applied parallel to the sample plane. When magnetic field applied parallel to the sample plane, the system is magnetically collinear and spin-flip reflectivity is equal to zero. Thus the neutron intensity in parallel applied field in Fig. 10 (closed symbols) is background due to imperfect polarizing efficiency of polarizer (mainly) and analyzer. In off-specular reflection region for the high applied field, in Fig. 10a one can see the maxima of the resonances n=3 and n=4. For the resonance n=3 the effect is equal to 0.03 count/s and background is equal to 0.005 count/sec. The ratio is *effect/background*=6.0. In specular reflection region for the low field in Fig. 10c one can see the resonances n=3 and 4. For the resonance n=3 the effect for the high field is equal to 0.10 count/s and background is 0.35. The ratio is *effect/background*=0.3. The same is for the resonance n=4. For the low field (Fig. 10c) we have the ratio *effect/background*=0.03/0.15=0.2. For the field (Fig. 10c) we have the ratio *effect/background*=0.012/0.003=4.0 what is in 20 times greater than for the low field. Thus, in off-specular reflection region the ratio *effect/background* is in 20 times greater than in the specular reflection region. In Fig. 10b one can see the resonances n=5 and n=6 for the state off,on for the high applied field. The resonances n=3 and n=4 are absent because of restriction due to Zeeman energy changing at spin-flip as for the reflected beam (-+) in Fig. 5c.

We demonstrated that in off-specular reflection region under the high applied field the spin-flip probability drops in 3 times but the background is reduced almost in $10^2$ times. Thus the ratio *effect/background* is increased in 20 times. In this case, the magnetic state of the saturated magnetic films is not changed. We only changed the quantization axis applying the high inclined magnetic field.

## VI. CONCLUSIONS

We considered experimentally the Zeeman spatial splitting of neutron beam at the reflection from the uniformly magnetized thin film. The high external magnetic field is applied perpendicular to the sample surface. The Zeeman beam-splitting effect is the separation in space the neutron beams of definite spin-flip transitions (-+) and (+-). It means that in theory in off-specular reflection region the neutron beam is perfectly polarized and the parasitic background (neutrons of another spin state) is absent. This property can be used for the construction of a simple polarizer with theoretical polarizing efficiency 100 %. The similar based on beam-splitting way to polarize monochromatic beam using a magnetic film placed in a high perpendicular field was proposed in [35]. In this communication we show experimentally that for the long neutron wavelength 8 - 10 Å the polarizing efficiency 0.97 is higher than typical polarization efficiency 0.8 at these long neutron wavelengths for supermirrors. In spite of the spin-flip reflectivity of the beam-splitting based polarizer is lower, the factor $P^2R$ is comparable with supermirrors for the long neutron wavelength. In particular cases the higher polarization degree and lower background is more important than higher neutron intensity at high background.

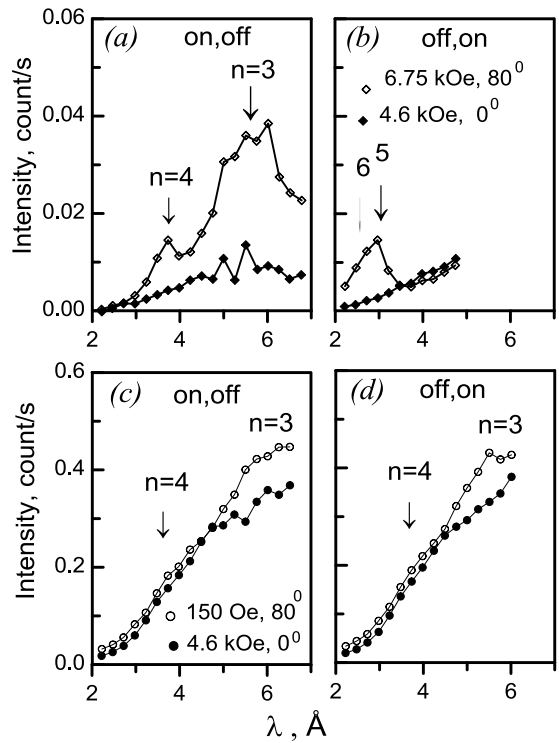

Fig. 10. Experimental neutron intensity for the sample Ti(30 nm)/Co(6)/Ti(200)/Cu(100)//glass(substrate) in off-specular *(a,b)* and specular *(c,d)* reflection regions as a function of neutron wavelength at the polarizing modes on,off *(a,c)* and off,on *(b,d)*. The external magnetic field 6.75 kOe *(a,b)* and 150 Oe *(c,d)* is applied under an angle 80° to the sample surface (open symbols). Closed symbols (background) correspond to the external field 4.6 kOe applied parallel to the sample surface. It is background.

Such particular case was demonstrated experimentally. Neutron standing waves in the planar waveguide with a magnetic layer were observed using neutron spin-flip in this magnetic layer. External magnetic field (high and low) was applied under an angle to the sample surface. In the high field at beam-splitting, the ratio *effect/background* is in 20 times greater than in specular reflection region in the low field due to background reducing. It is the model experiment for demonstration with the same magnetic state of the layer but different quantization axis of the neutron spin.

It is possible to observe Zeeman beam-splitting not only in high perpendicular applied field which may change the magnetic state of the magnetic layer. We can also register beam-splitting in low parallel external field if there are magnetically noncollinear regions with high internal magnetic field, for example magnetic domains [11] or clusters [16].

### Acknowledgements

The authors are thankful to V.L. Aksenov and Yu.V. Nikitenko for fruitful discussions.